\documentclass[prl,aps,twocolumn,floats,superscriptaddress]{revtex4}
\usepackage{graphicx}
\usepackage{amssymb}
\usepackage{times}

\def\non{\nonumber}

\def\be{\begin{equation}}
\def\ee{\end{equation}}
\def\bea{\begin{eqnarray}}
\def\eea{\end{eqnarray}}
\newcommand{\bse}{\begin{subequations}}
\newcommand{\ese}{\end{subequations}}

\newcommand{\eqref}[1]{(\ref{#1})}

\def\r{{\bf r}}
\def\k{{\bf k}}
\def\q{{\bf q}}

\def\G{{\cal G}}
\def\L{{\cal L}}

\def\lp{\left(}
\def\rp{\right)}

\def\la{\left<}
\def\ra{\right>}

\def\w{\omega}

\def\Re{\mbox{\,Re\,}}
\def\Im{\mbox{\,Im\,}}
\def\Gbare{{G^{0}}}
\def\R{{\cal R}}
\def\half{\textstyle{\frac12}}

\begin{document}

\title{Transverse NMR relaxation in magnetically heterogeneous media}

\author{D.~S.~Novikov}
\email{dima@alum.mit.edu}
\affiliation{Department of Physics, Yale University, New Haven, CT 06520, USA} 
\author{V.~G.~Kiselev}
\email{kiselev@ukl.uni-freiburg.de} 
\affiliation{Department of Radiology, Medical Physics, 
University Hospital Freiburg, Hugstetterstr. 55, D-79106 Freiburg, Germany}
\date{\today}


\begin{abstract}
\noindent
We consider the NMR signal from a permeable medium
with a heterogeneous Larmor frequency component that
varies on a scale comparable to the 
spin-carrier diffusion length.
We focus on the mesoscopic part of the transverse relaxation,
that occurs due to dispersion of precession
phases of spins accumulated during diffusive motion.
By relating the spectral lineshape to correlation functions 
of the spatially varying Larmor frequency, we demonstrate how 
the correlation length and the variance of the Larmor frequency distribution
can be determined from the NMR spectrum. We corroborate our results 
by numerical simulations, and apply them to quantify human blood spectra.
\end{abstract}




\maketitle


\section{Introduction}
\label{sec:intro}

Transverse relaxation of the NMR or ESR signal 
acquires distinctive 
features \cite{Glasel74} when it occurs in a medium which possesses magnetic structure on a {\it mesoscopic scale}.  
This scale is intermediate between the microscopic atomic or molecular scale,
and the macroscopic sample size or 
the resolution achievable with time-resolved Faraday
rotation measurement \cite{AwschalomBook2002},
magnetic resonance imaging  \cite{VlaardingerbroekBook} (MRI), or ESR imaging 
\cite{EatonBook91}.
In this work we consider the transverse relaxation 
in a broad class of media where the mesoscopic structure 
can be characterized by variable Larmor frequency $\Omega(\r)$. 

The problem of mesoscopic contribution to transverse relaxation
arises in a broad variety of contexts, 
ranging from solid state physics to radiology. 
In the semiconductor spintronics, the 
spatially dependent Larmor frequency $\Omega(\r)$ for electrons or holes
can be induced either via ferromagnetic imprinting,
or electrostatically by locally varying the electron $g$-factor 
\cite{Crooker97,Kikkawa97,Kikkawa98,Salis2001,Lau2006}. 
For nuclear or electron spins in liquids, the spatially dependent Larmor frequency 
can arise due to heterogeneous magnetic susceptibility.
The latter property is crucial in the field of biomedical MRI, where 
the heterogeneous susceptibility $\chi(\r)$ is inherent to
majority of living tissues due to paramagnetism of 
deoxygenated haemoglobin in red blood cells \cite{Ogawa90_bold1,Ogawa90_bold2},
enabling {\it in vivo} visualization of regional activations in human brain
via functional MRI \cite{Belliveau91,Kwong92,Bandettini92,Menon92}. 
Moreover, the susceptibility contrast 
can be enhanced by doping blood with magnetic contrast agents 
for clinical purposes, e.g. for diagnostics of acute stroke.

The common property of all these  
systems is the presence of 
the transverse relaxation that occurs via the dispersion of precession
phases of spins accumulated during their diffusive motion.
This reduces the vector sum of magnetic moments from all spins in the sample
and causes attenuation (termed {\it dephasing}) 
of the measured signal time course $s(t)$.
The signal $s(t)$ is typically a sum over a large number of spins
acquired over a macroscopic volume $V$ whose size greatly exceeds 
the diffusion length $l_D\propto \sqrt{t}$. 
The randomness of the Brownian trajectories results in an effective averaging 
(diffusion narrowing) of the contributions
from different parts of the volume $V$ that are separated by 
less than $l_D$. 
Further averaging occurs between larger domains of size exceeding $l_D$.
This self-averaging character of the measurement is a major obstacle in 
quantifying the properties of the medium on the scales below $V$.
Here we study how the {geometric} structural details of 
$\Omega(\r)$ on the mesoscopic scale are reflected in the measured signal,
i.e. survive the above averaging.
For definitiveness, we will use the established NMR terminology,
since generalizations onto the ESR case present no problem.

Theory of mesoscopic relaxation in the presence of the diffusion narrowing 
has been previously addressed in the MRI context 
\cite{Gillis87,Kennan94,Kiselev98,Jensen2000_dnr,Bauer99,Bauer2002,Kiselev2002,Sukstanskii2003,Sukstanskii2004_dnr}.
These works, however, either do not directly relate 
relaxation to the structure \cite{Kennan94}, 
or employ strong simplifying assumptions 
(representing the medium as a dilute suspension of mesoscopic objects with 
small volume fraction $\zeta\ll 1$ 
\cite{Gillis87,Kiselev98,Jensen2000_dnr,Bauer99,Bauer2002,Kiselev2002,Sukstanskii2003,Sukstanskii2004_dnr} 
and a particular shape
\cite{Gillis87,Kiselev98,Jensen2000_dnr,Bauer99,Bauer2002,Sukstanskii2003,Sukstanskii2004_dnr}). 
This leads 
to the virial expansion for the signal $s(t) \propto  e^{-\zeta f(t)}$, where
$f(t)$ describes the dephasing due to a single object.
The mesoscopic field heterogeneity is accounted 
for perturbatively, which yields relaxation $f \propto \Omega^2$.
As a consequence, the known results in the diffusion-narrowing regime 
are limited to dilute suspensions of effectively weakly magnetized objects. 
Examples are dilute blood samples, 
or tissues with small volume fraction of paramagnetic vessels,
in the fields $B_0 \lesssim 1\,$T.

The aim of this paper is to provide a framework for the transverse 
relaxation from arbitrary magnetic media beyond the limitations 
of dilute suspension and the weak magnetization. 
First, we suggest a universal description of the spectral lineshape $s(\omega)$ 
of the signal:
\be \label{s-omega}
s(\omega) = {1\over -i\omega - \Sigma(\omega)} \,, 
\ee
where function $\Sigma(\omega)$ is a measurable characteristic of the medium.
The quantity $-\Sigma(\omega)$ can be loosely interpreted as a dispersive relaxation rate, with
Eq.~(\ref{s-omega}) being a generalization of the conventional Lorentzian line shape to the case of heterogeneous magnetic medium.
Technically, the representation (\ref{s-omega}) originates from the standard form of single particle Green's function in many-body physics, with $\Sigma(\omega)$ called the {\it self-energy} part \cite{AGD}, the term we will use here.
The immediate advantage of using $\Sigma(\omega)$ instead of the traditional 
lineshape $s(\omega)$ is that the frequency-dependent part of the self-energy 
is entirely determined by the mesoscopic magnetic structure and thus 
vanishes for homogeneous media. In contrast, the effect of mesoscopic structure 
on the lineshape $s(\omega)$ is a complicated deformation of a Lorentzian, 
that is difficult to quantify \cite{Kuchel89,Bjoernerud2000}. 
Further we relate the self-energy to the geometric structure on the mesoscopic scale for media fully permeable for spin carriers,
and verify the results using {\it ab intio} 
simulations of the transverse relaxation. 
Finally, we apply our general results 
to quantify the line shape of water proton resonance in blood \cite{Bjoernerud2000} 
in terms of mesoscopic structural parameters.

\section{Results}
\label{sec:results}

Here we consider the NMR signal from a {\it random medium} 
which is characterized by the correlation functions 
$\Gamma_n(\r_1, ... , \r_n) = \la \Omega(\r_1) ... \Omega(\r_n)\ra$.
These functions vary on the mesoscopic scale much smaller than the size of  
the acquisition volume $V$.
The self-averaging character of the measurement implies that one needs to find
the signal from a particular realization of $\Omega(\r)$, and then 
to average over the realizations according to the distribution moments $\Gamma_n$.
In what follows, we suppose for simplicity that the medium is isotropic and 
translation-invariant in a statistical sense. These assumptions cover 
a wide variety of applications, such as the transverse relaxation
in blood, or in the brain gray matter.

\subsection{Spectral lineshape}
\label{sec:BT}

A spin traveling along Brownian path $\r(t)$ originating at $\r(0)=\r_0$,
acquires the relative phase $\exp\left\{-i\int_0^t\! dt\, \Omega[\r(t)]\right\}$.
Here $\Omega(\r)$ is the variable component of the Larmor frequency; 
$\la \Omega(\r)\ra \equiv 0$. 
The time evolution 
\be \label{PI}
\G(\r_0,\r; t) = \int_{\r_0}^\r {\cal D}\r(t) \, 
e^{-i\int_0^t\! dt\, \Omega[\r(t)] - \int_0^t\! dt\, \dot\r^2/4D}
\ee
of the magnetization packet initially at $\r=\r_0$,
$\G(\r_0,\r, t)|_{t=0} = \delta(\r-\r_0)$,
is the precession phase
averaged over the Wiener measure on the diffusive paths. 
Equivalently, $\G(\r_0,\r; t)$
is the {\it Green's function} (fundamental solution) of
the Bloch-Torrey equation \cite{Torrey56}
\be \label{BT}
\partial_t \psi
= \nabla_\r (D\nabla_\r \psi)  - i\Omega(\r)  \psi \,.
\ee
The diffusive dynamics \eqref{BT} leads to the transverse relaxation,
$\G(\r_0,\r, t)|_{t\to\infty} = 0$.

For simplicity, and in order to underscore the effects of susceptibility contrast,
below we assume homogeneous diffusivity $D(\r)\equiv D={\rm const}$. 
We also excluded conventional exponential factors associated with
homogeneous components of Larmor frequency and of transverse relaxation.

The mesoscopic component $s(t)$ of the 
measured MR signal (normalized to $s|_{t=0}\equiv 1$) is obtained in 
two stages. First, one 
considers the signal $s_\Omega$ from a given realization of $\Omega(\r)$.
It is given by the magnetization $\G(\r_0,\r; t)$
that is summed over the final positions $\r$ which spins reach during the time interval $t$, and averaged over initial spin positions $\r_0$.
The signal $s(t)$ is then found by averaging of $s_\Omega$ 
over all realizations of $\Omega(\r)$, 
\be \label{s-G}
s(t) = \la s_\Omega(t)\ra 
= \int \! {d\r d\r_0 \over V} \, \la \G(\r_0, \r; t)\ra  \equiv 
G(t;\k)|_{\k=0} \,.
\ee
Here we used the translation invariance of the 
distribution-averaged Green's function $\la \G(\r_0,\r; t)\ra  \equiv G(\r-\r_0, t)$,
with its Fourier components defined as
$G_{\omega,\k} = \int\! dt d\r \, G(\r,t) e^{i\omega t - i\k\r}$.

Eq.~\eqref{s-G} connects the spectral lineshape $s(\omega) = G_{\omega,\k}|_{\k=0}$
with the distribution-averaged Green's function of Eq.~\eqref{BT}.
The result of such an averaging (see {\it Methods})
may be represented as
\be \label{G}
G^{-1}_{\omega,\k} = \Gbare_{\omega,\k}^{-1} -\Sigma_{\omega,\k} \,,
\ee
where the diffusion propagator
\be \label{Gbare}
\Gbare_{\omega,\k} = {1 \over -i\omega + D k^2}  
\ee
is the Green's function of Eq.~\eqref{BT} with $\Omega\equiv 0$.

The generic representation \eqref{G} yields the dynamics $\omega=\omega(\k)$
as the pole of the propagator $G_{\omega,\k}$,
modified by interaction with complex environment embodied in $\Sigma_{\omega,\k}$
\cite{AGD}. 
For the ballistic dynamics, 
expansion of $\Sigma_{\omega,\k}$ would give the refraction index of the medium
(when $\omega \propto k$), or the renormalized quasiparticle mass and lifetime
(when $\omega \propto k^2$). In the present case of the diffusive 
dynamics \eqref{BT} ($i\omega\propto k^2$), 
the {self-energy} $\Sigma_{\omega,\k}$ 
contains all the measurable information about mesoscopic relaxation. 
In particular, its expansion 
$\Sigma_{\omega,\k} = \Sigma(0) - (\delta D) k^2 + ...$ in $\omega$ and $\k$ 
describes the effect of the mesoscopic relaxation
on the coarse-grained dynamics of the magnetization density.
In other words, Eq.~\eqref{G}, viewed as an operator relation, is the effective 
Bloch-Torrey equation, that acquires higher derivatives in $t$ and $\r$, 
after averaging over the medium.
According to Eq.~\eqref{s-G}, the signal lineshape \eqref{s-omega} 
is a particular case of \eqref{G} with
\be 
\Sigma(\omega) \equiv \Sigma_{\omega; \k}|_{\k=0} \,. 
\ee

\subsection{Weak dephasing: A perturbative solution for the lineshape}
\label{sec:2nd}

As a simplest example we now consider the weakly magnetic medium 
whose Larmor frequency dispersion $\delta\Omega\equiv \sqrt{\la \Omega^2 \ra}$ 
is small, and find the relaxation in the lowest order in 
$(\delta\Omega)^2 \equiv \Gamma_2(\r)|_{\r=0}$.
The self-energy is given by the single Feynman graph 
(see {\it Methods}, Fig.~\ref{fig:diag}(c)),
\be \label{sigma2-wk}
-\Sigma^{\rm pert}_{\omega,\k} = \int \! {d^d\q \over (2\pi)^d}\, 
{\Gamma_2(\q) \over -i\omega + D (\k+\q)^2} \,.
\ee
The lowest order contribution to the self-energy part for the signal,
\be \label{sigma2}
-\Sigma^{\rm pert}(\w) = \int \! {d^d\q \over (2\pi)^d}\, 
{\overline{\Gamma_2}(q) \over -i\omega + D q^2} 
\ee
involves only the angular-averaged two-point correlation function 
$\overline{\Gamma_2}(q)=\la \Gamma_2(\q)\ra_{\hat \q}$.
Eq.~\eqref{sigma2} is the Gaussian approximation in a weakly paramagnetic medium. 
Corrections to this approximation, involving correlators $\Gamma_n$,
first appear in the $n$-th order in the Born series for $\Sigma_{\w,\k}$.

Equation \eqref{sigma2} gives the universal short-time asymptotic behavior:
In the limit $\omega\to \infty$ this is the leading in $1/\omega$
contribution to $\Sigma(\w)$, yielding
$\Sigma(\w)|_{\w\to\infty} \simeq (\delta\Omega)^2/i\w$. 
Substituting it into Eq.~\eqref{s-omega} and taking the inverse Fourier transform, 
results in $s(t)|_{t\to 0} \simeq 1 -  (\delta \Omega \cdot t)^2/2$, for 
$\delta\Omega \cdot t \ll 1$.

\begin{figure}[t]
{\bf (a)}\\
\includegraphics[width=40mm]{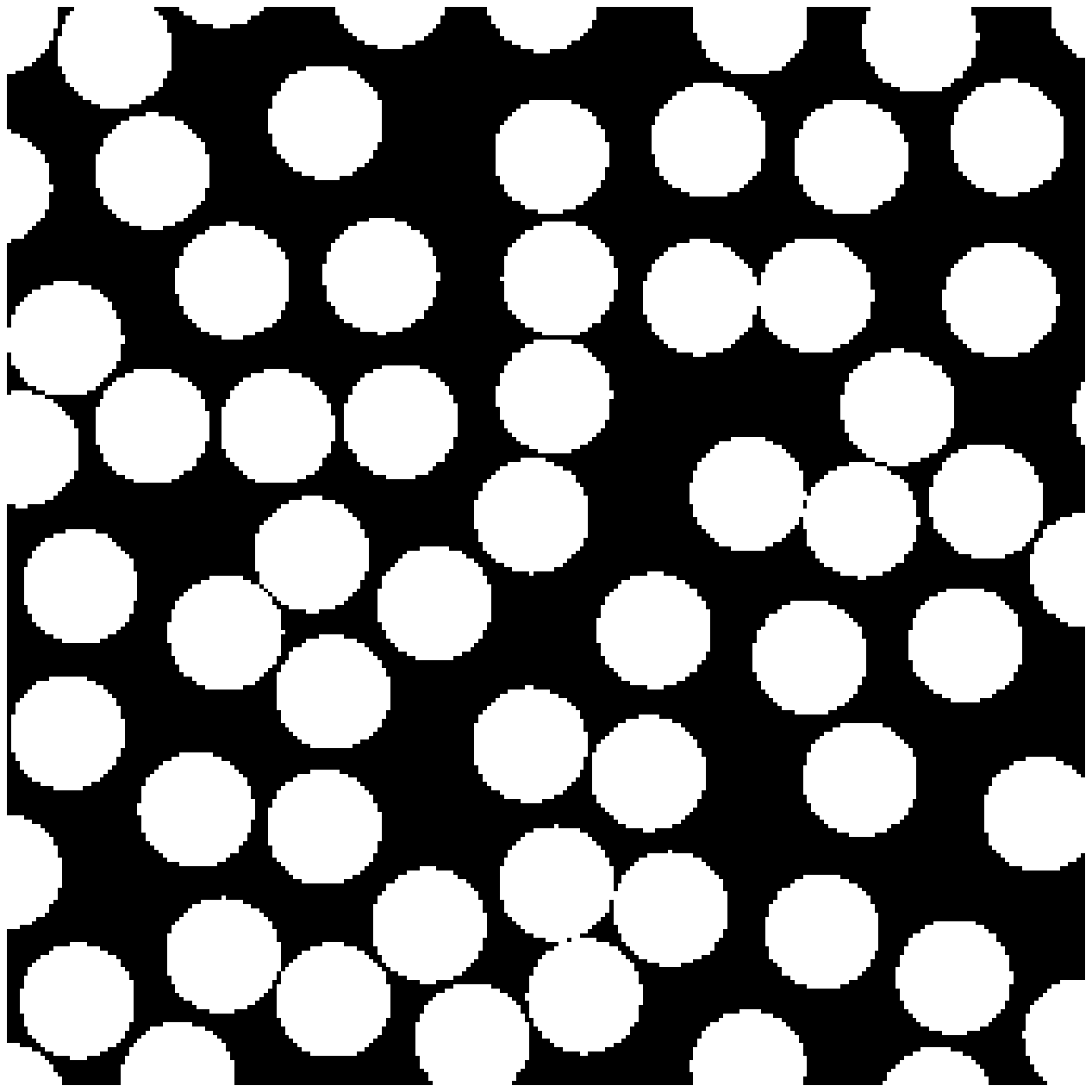}
\includegraphics[width=40mm]{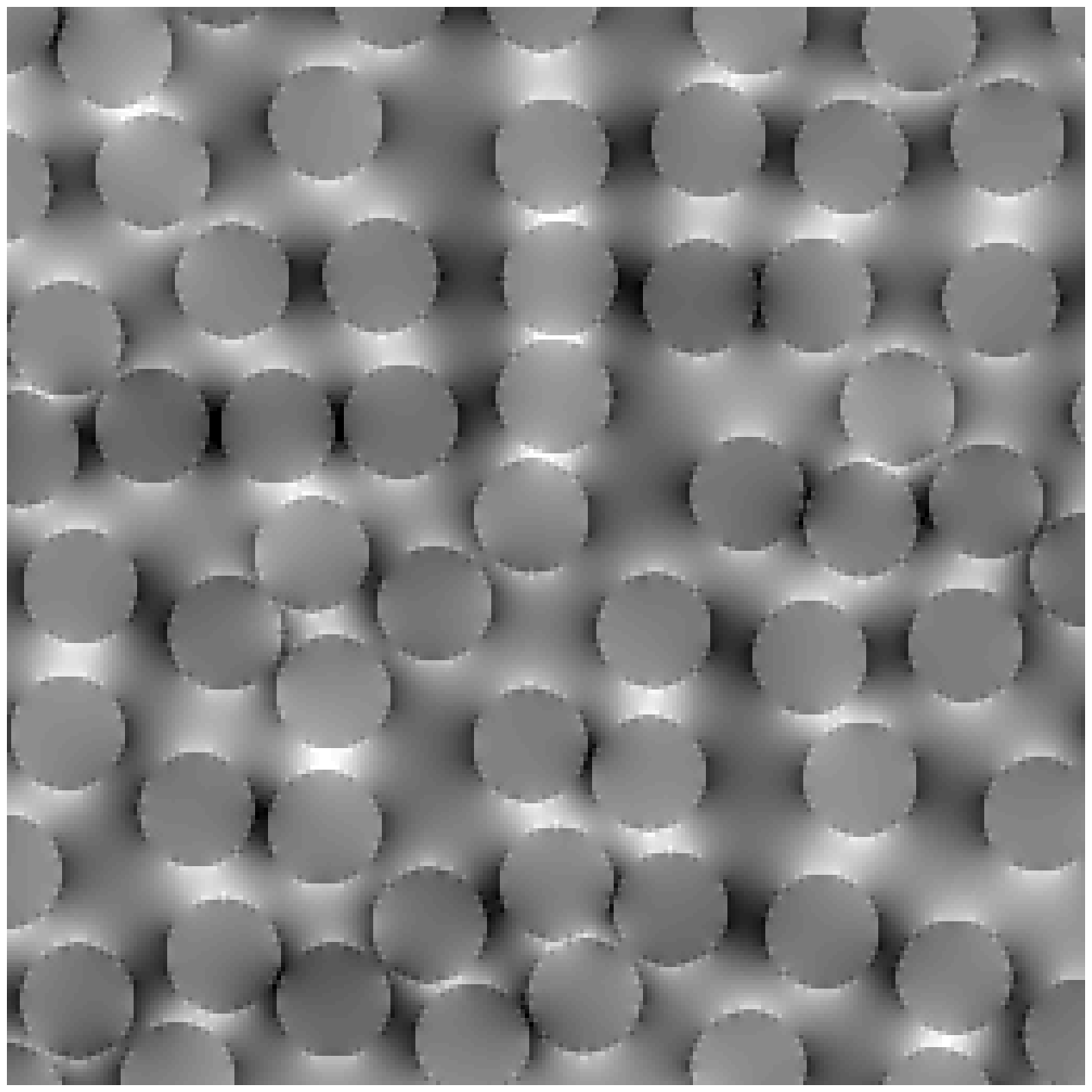}
\\[2pt]{\bf (b)}\\
\includegraphics[width=40mm]{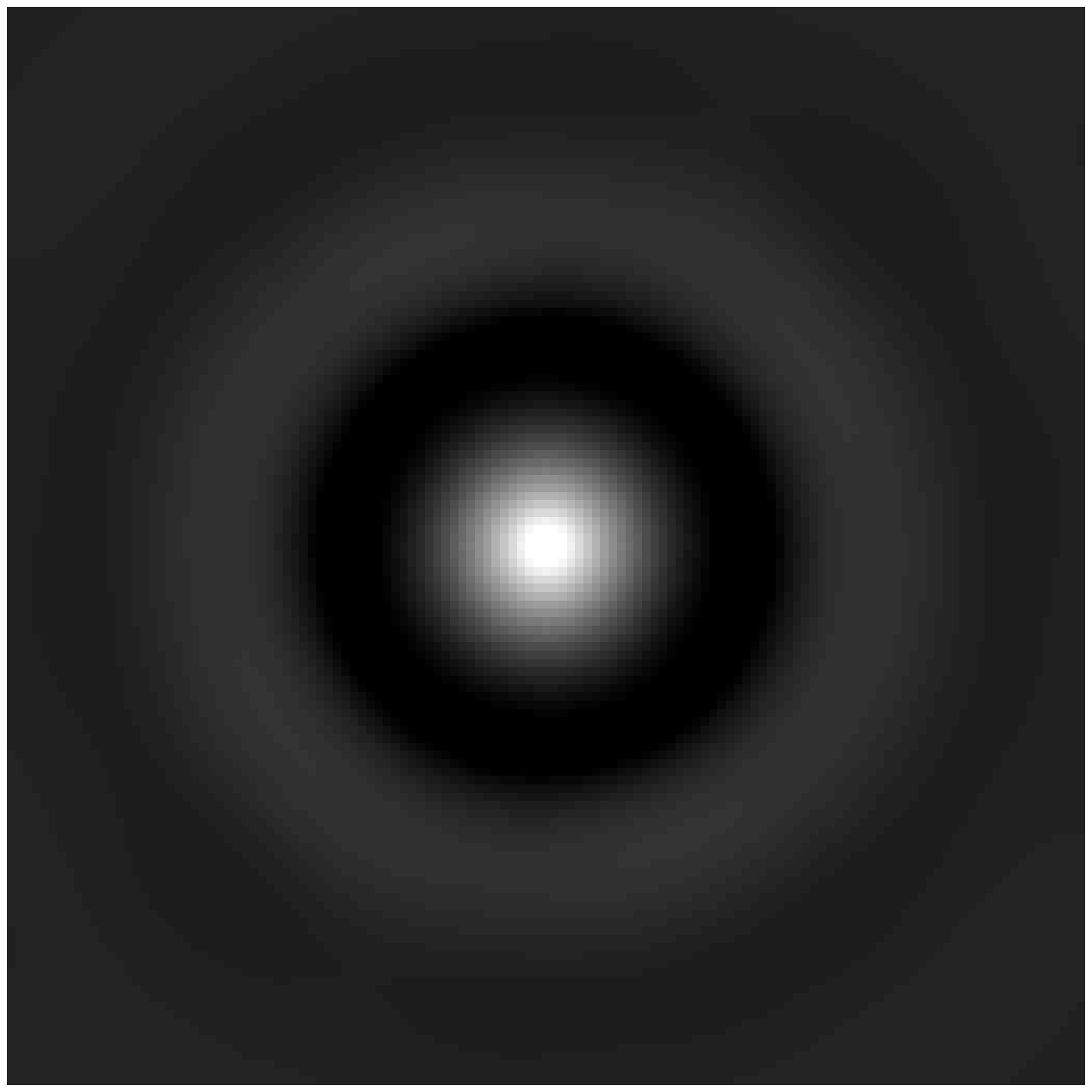}
\includegraphics[width=40mm]{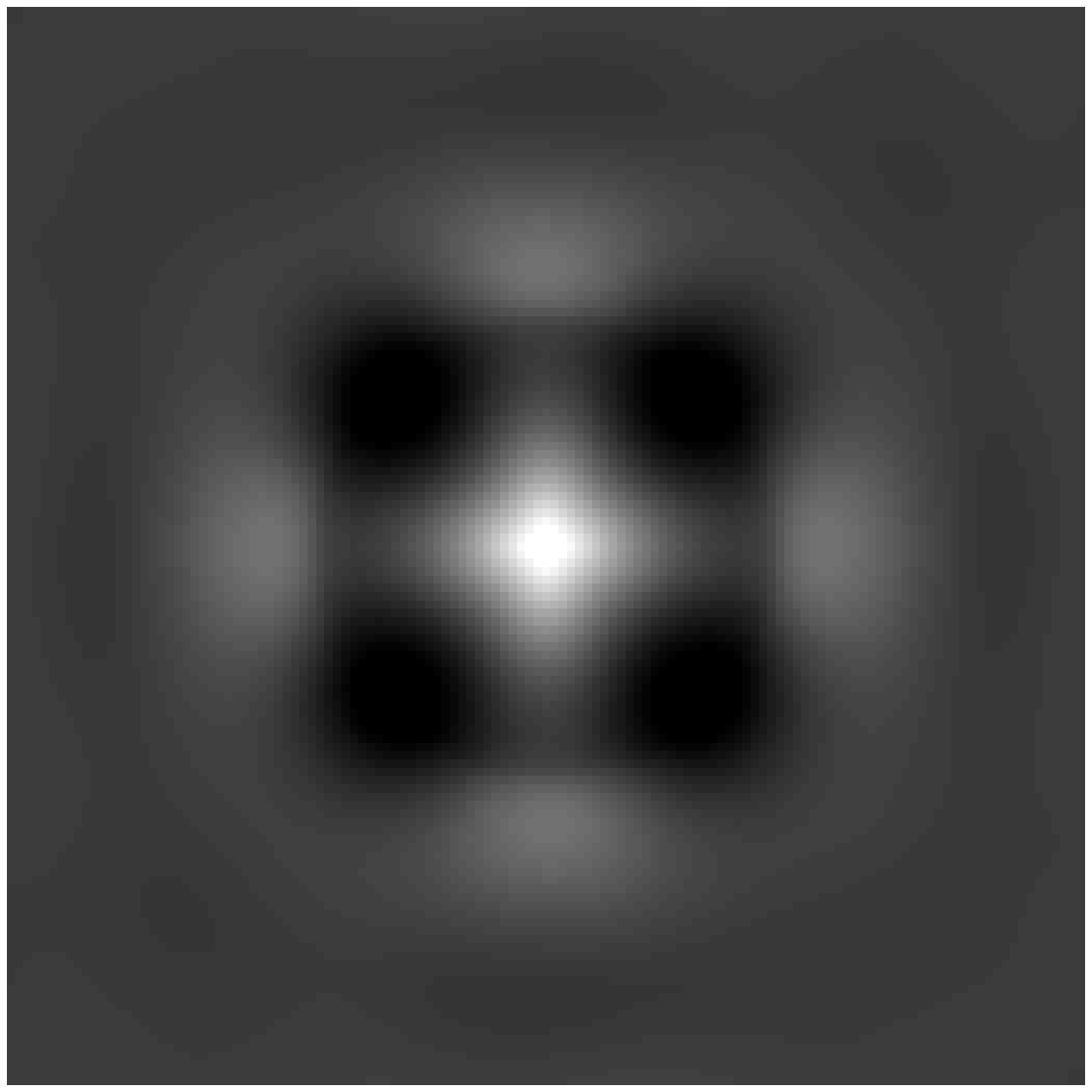}
\\[2pt]{\bf (c)}\\
\includegraphics[width=42mm]{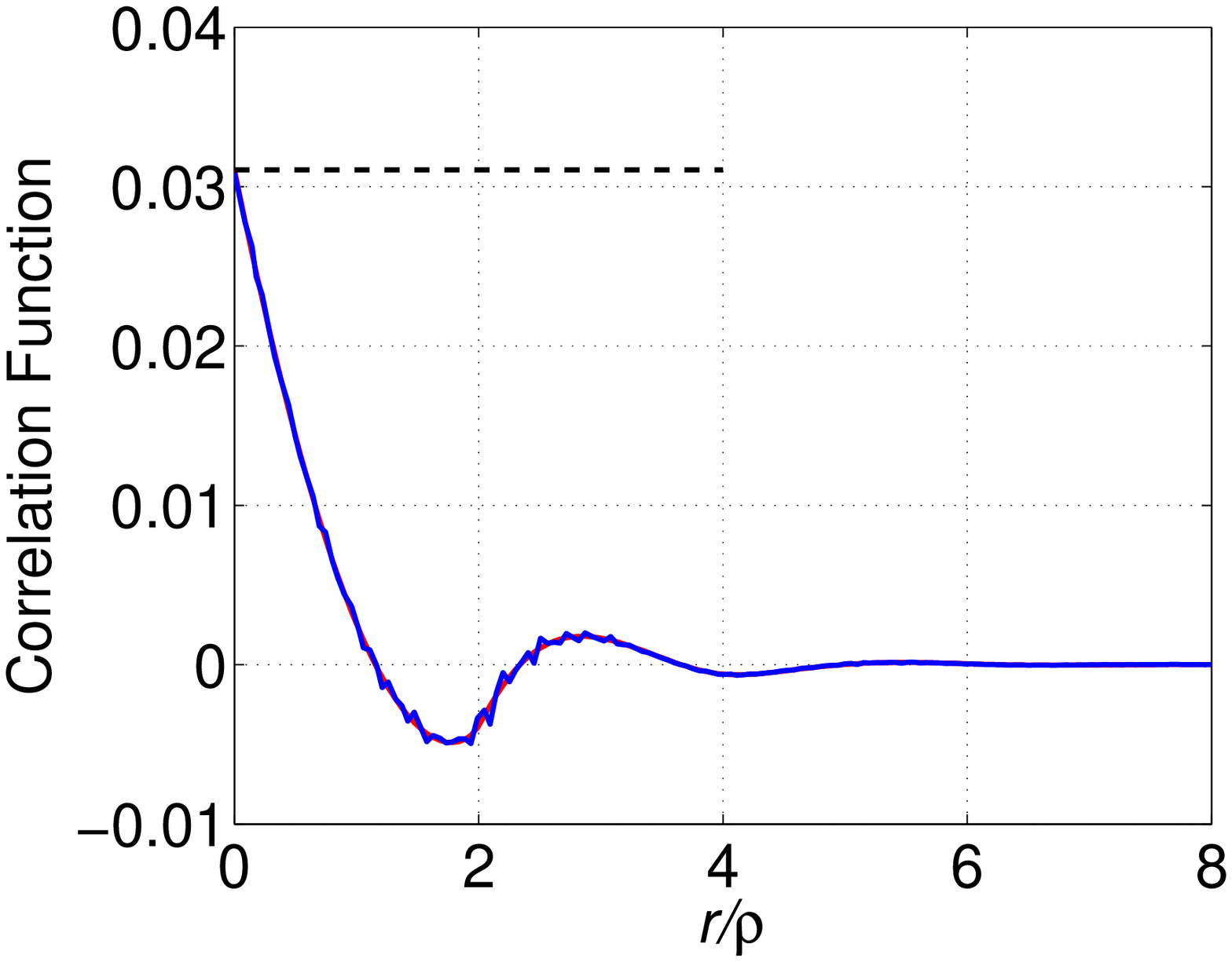}
\includegraphics[width=42mm]{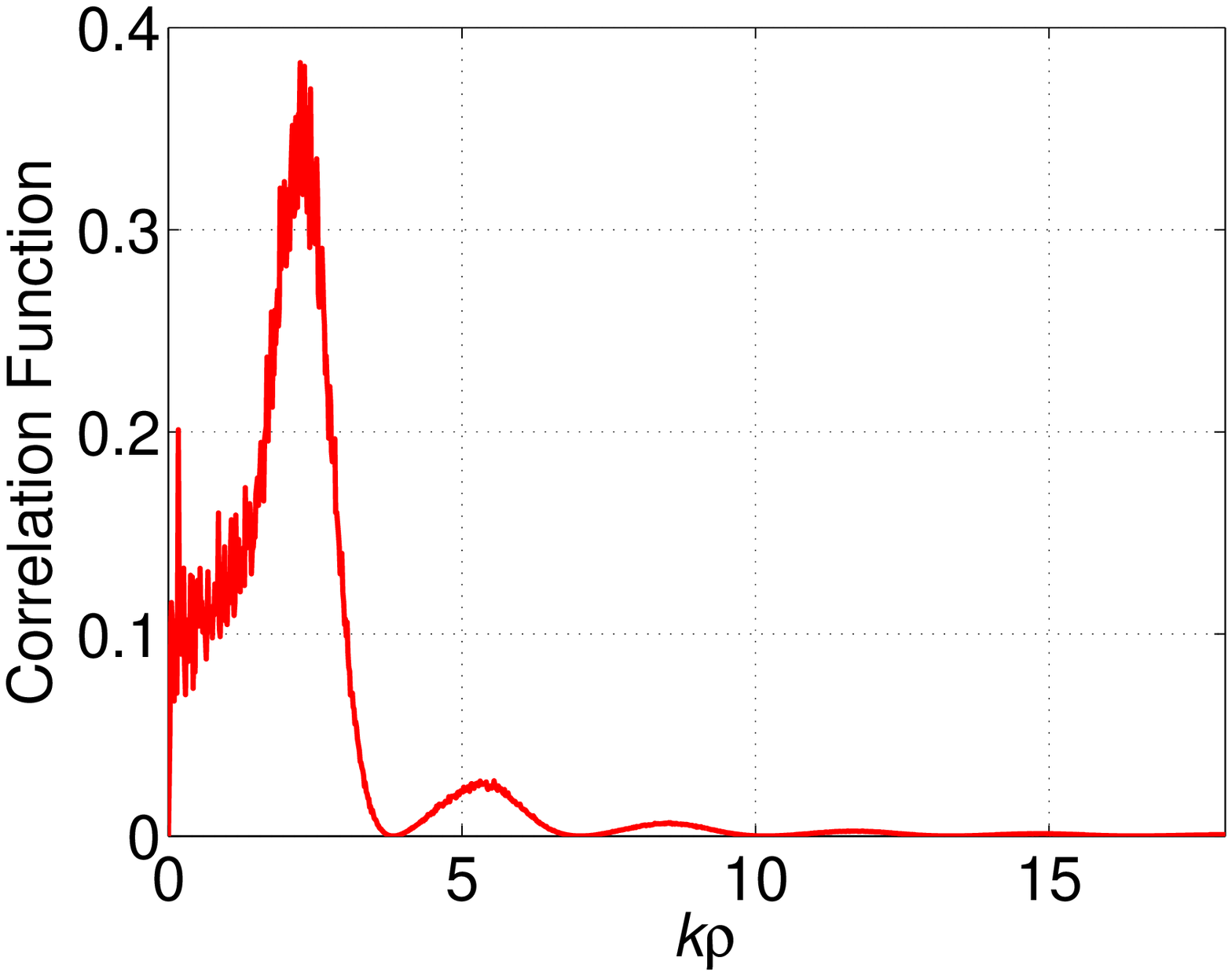}
\caption{\footnotesize
The two-dimensional medium. 
(a) The susceptibility profile $\chi(\r)$ 
generated by the random self-avoiding addition of disks of radius $\rho$
with volume fraction $\zeta=0.461$,
and the corresponding Larmor frequency offset $\Omega(\r)$
induced by a vertically applied field. The shown fragment is 16 times 
smaller in each dimension than the whole simulation box of size $300\rho$.  
The diffusion of spins was modeled as random hopping on a square lattice 
($4096\times 4096$) with lattice constant $300\rho/4096$, 
in the field shown in right panel.
Time was measured in units of $t_\rho=\rho^2/D$. 
(b) 
Correlation functions $\Gamma_2^\chi(\r)$ and $\Gamma_2(\r)$
of the susceptibility and of the Larmor frequency. 
The images are zoomed 4-fold as compared with (a).  
(c) 
Illustration of locality in $d=2$. Left panel: Coinciding
angular-averaged correlators of the susceptibility $\Gamma_2^\chi(r)$ 
(red) rescaled by the factor $\lambda_{2d}^2=1/8$, and of 
the Larmor frequency (blue). Dashed horizontal line corresponds 
to expected value $\zeta(1-\zeta)/8$ at $r=0$. 
Right panel shows the angular-averaged Fourier transform of the 
correlator, $\overline{\Gamma_2}(k)$, 
with the pronounced peak at $k_c \approx 2.3/\rho$.
Noise increases for small $k$ due to finite size effects; $\Gamma_2|_{k=0}=0$.  
The mean $k$ of the first peak found with the weight $k$, 
which is inherent to the 2d integration, is $k_c=2.20/\rho$. 
}
\label{fig:2d}
\end{figure}

\subsection{Connection to the susceptibility structure: Locality}

The diffusing spins sense the distribution of the 
Larmor frequency offset $\Omega(\r)$. Often times, in paramagnetic media, 
such a distribution is induced by a variable 
magnetic susceptibility profile $\chi(\r)$, which is practically interesting.
Then the medium is characterized by the correlation functions of $\chi$, 
assuming zero average $\la\chi\ra=0$. In particular, here we consider
the two-point correlation function $\Gamma_2^\chi(r)=\la \chi(\r)\chi(0)\ra$
that for a statistically isotropic medium depends only on $r=|\r|$.
For nonferromagnetic media, such as deoxygenated blood,
$|\chi|\ll 1$, hence the induced field $\Omega(\r)$ is connected to $\chi(\r)$
by the convolution 
with an  elementary dipole field $Y(\r)$, 
\be \label{Y-r}
\Omega(\r)=4\pi \Omega_0 Y(\r)\!*\!\chi(\r)\,, \quad 
Y(\r) = {1 \over 4\pi r^3} \lp {3z^2 \over r^2} - 1 \rp .
\ee
Here $\Omega_0$ is the uniform Larmor frequency component,
and $Y(\r)$
includes both the local contribution 
$\propto \delta(\r)$, and the Lorentz cavity
field, that compensate each other in three  
dimensions \cite{JacksonBook,Dickinson51}.

The angular-averaged correlators of Larmor frequency and of the underlying
susceptibility are proportional to each other ({\it locality} \cite{Kiselev2002}): 
\bea \label{locality}
\overline{\Gamma_2}(r) &=& \la \Omega(\r)\Omega(0)\ra_{\hat\r}
 = (4\pi \lambda \Omega_0 )^2 \cdot \Gamma_2^\chi(r) \,, 
\\
\label{def-lambda}
\mbox{where}\ \ \lambda^2 
&=& \int \! d\hat\k \, Y(-\hat\k)Y(\hat \k) \,. 
\eea
Here $Y(\hat\k)$ is the Fourier transform of the dipole field;  
in $d=3$ dimensions, $Y(\hat \k) = 1/3-{k_z^2/k^2}$,
and Eq.~\eqref{def-lambda} yields $\lambda^2=4/{45}$.
In $d=2$ dimensions, $Y(\hat \k) = 1/2-{k_y^2/k^2}$, yielding $\lambda^2=1/8$.

To prove Eq.~\eqref{locality},
we note that the Fourier transform of 
the field $\Omega$ induced by the susceptibility profile 
with the orientation $\R$ is
$\Omega_\R(\k) = 4\pi\Omega_0 Y(\hat\k) \chi(\R^{-1}\k)$.
Substituting $\Omega_\R$ into first Eq.~\eqref{locality}
and averaging over orientations $\R$ yields the right-hand side.
Physically,
we went from averaging over the orientations $\R$ of the medium relative to a given 
$\hat {\bf z}$-direction of the main field $B_0 \hat{\bf z}$ 
to averaging over the direction of the field at fixed 
orientation of the structure $\chi(\r)$.

The locality property \eqref{locality} 
means that, in the second order in $\Omega$, after the orientational averaging 
the diffusing spins effectively interact {\it directly} 
with the susceptibility profile $\chi(\r)$, via $4\pi\lambda\Omega_0\chi(\r)$. 
This interaction is much simpler 
than that with the susceptibility-induced field 
$\Omega(\r)$, which involves a convolution with the non-local field \eqref{Y-r}
with a complicated angular dependence. The underlying reason for 
this fortunate property is the scaling $Y(\r) \propto r^{-d}$ in $d$ dimensions
\cite{Kiselev2002}.
The locality \eqref{locality} is demonstrated for the two-dimensional 
numerically generated medium in Fig.~\ref{fig:2d}.

\subsection{Model of magnetic structure}

Aimed at relating the signal to the magnetic structure,
we consider the medium, embodied by the distribution of $\chi(\r)$, 
that has a single mesoscopic length scale $l_c$.
This scale, and the dispersion $\delta\Omega$, 
are the two parameters that we employ here to characterize 
the medium.

For the model calculations below we focus on the angular-averaged correlation 
function $\overline{\Gamma_2}$ entering Eq.~\eqref{sigma2}.  
The presence of a single length scale leads to a well-defined dominant peak 
in the function $\overline{\Gamma_2}(k)$.
Such a form is demonstrated for the $d=2$ case in Fig.~\ref{fig:2d}(c).
We suggest to approximate such a peak via a delta-function
\be \label{Gamma-delta}
d=3: \quad \overline{\Gamma_2}(k) = \half (\delta\Omega)^2 l_c^2 \delta(k-k_c) \,, 
\quad k_c = 2\pi/l_c \,,
\ee
normalized to 
$\int\! {d^d\k\over (2\pi)^d} \, \overline{\Gamma_2} \equiv (\delta\Omega)^2$,
assuming that the contribution of other length scales is less relevant.
In real space, the correlator \eqref{Gamma-delta} 
\be \label{Gamma-delta-r}
d=3: \quad \overline{\Gamma_2}(r) = (\delta\Omega)^2 \, {\sin k_c r \over k_c r} \,. 
\ee
In $d=2$ dimensions, the corresponding correlator is
$\overline{\Gamma_2}(k) = (\delta\Omega)^2 l_c \delta(k-k_c)$ 
and $\overline{\Gamma_2}(r) = (\delta\Omega)^2 J_0(k_c r)$, 
where $J_0$ is the Bessel function.
Generalization onto the case when the medium has a set of harmonics 
\eqref{Gamma-delta-r} with different $k_c$ is straightforward: 
e.g. for $d=3$,
\be \label{Gamma-sum}
\overline{\Gamma_2}(k) 
= \frac12 (\delta\Omega)^2 \sum_j p_j {l_c}_j^2 \delta(k-{k_c}_j) 
\quad \mbox{with}\quad \sum_j p_j =1 \,.
\ee

With the approximation \eqref{Gamma-delta},
evaluation of self-energy \eqref{sigma2} in the lowest order is trivial:
\be \label{Sigma2}
- \Sigma^{\rm pert}(\w) = {\alpha^2/t_c \over 1 - i\w t_c} \,.
\ee
Here we introduced the diffusion time
\be \label{tc}
t_c = {1/Dk_c^2} = {(l_c/2\pi)^2 /D} 
\ee
past the Larmor frequency correlation length $l_c=2\pi/k_c$. Time $t_c$  
sets the scale for both the frequency $\w$ and for the self-energy.
The  dimensionless perturbation series parameter
\be \label{alpha}
\alpha  \equiv 
\delta \Omega \cdot t_c 
\ee
controls the dephasing strength. 

The lineshape \eqref{s-omega}, \eqref{Sigma2} 
is a sum of the two Lorentzians
\bea \label{s-pert}
s(\w) &=& {w_1\over \epsilon_1 -i\w} + {w_2\over \epsilon_2 -i\w} \,,
\quad w_1+w_2=1\,, \\
s(t) &=& w_1 e^{-\epsilon_1 t} + w_2 e^{-\epsilon_2 t}
\label{s-pert-t}
\eea
where (keeping terms up to the order $\alpha^2$)
\be \label{epsilon12-pert}
\epsilon_1 t_c = 1-\alpha^2 \,, \quad \epsilon_2 t_c = \alpha^2 \,,
\ee
and the weights $w_1 = -\alpha^2$ and $w_2=1+\alpha^2$.
The approach \eqref{sigma2} is valid for $\alpha\ll 1$. 
The signal 
$s(t)  
 \simeq 1 - \half\alpha^2 (t/t_c)^2$ for $t\ll t_c$,
and decays monoexponentially for $t\to \infty$ as 
$s \simeq w_2 \exp\{-t/T_2^{\rm pert}\}$ with the rate
$1/T_2^{\rm pert} = \alpha^2/t_c = (\delta\Omega)^2 t_c$.
The latter rate arises due to angular diffusion of the precession phase:
As the phase changes by $\sim \alpha$ at each time step $t_c$, 
over large time $t\gg t_c$ the rms phase $\sim \alpha \sqrt{t/t_c}$ reaches
$\sim 1$ for $t\sim T_2^{\rm pert}$
(Dyakonov-Perel relaxation \cite{Dyakonov71}). 
The relaxation rate decreases with decreasing
$l_c$, as the medium effectively becomes more homogeneous 
(diffusion narrowing).

\subsection{Strong structural dependence}

What happens in a complex medium with a few harmonics, Eq.~\eqref{Gamma-sum}? 
In the lowest order, ${\cal O}\lp (\delta\Omega)^2\rp$, 
harmonics with different ${l_c}_j$ 
and dephasing strengths $\alpha_j =  \sqrt{p_j} \,\delta\Omega \cdot {t_c}_j$
contribute to the self-energy \eqref{sigma2} additively. Thus $\Sigma(\omega)$ is
a sum of $n$ terms \eqref{Sigma2}, and the lineshape is 
a sum of $n+1$ Lorentzians that correspond to the presence of $n$
distinct length scales. 

We emphasize that the apparent $n+1$-exponential form of the perturbative
result $s(t)$ (biexponential \eqref{s-pert-t} for $n=1$) 
here has nothing to do with existence 
of $n+1$ macroscopic ``compartments'' with different relaxation 
rates. (Note also that the weight $w_1<0$ in Eq.~(\ref{s-pert-t}).) 
This simple example is a warning
against common practice of literally interpreting bi- or 
multi-exponential  fits.

The $t\to\infty$ relaxation
is determined by the pole of signal \eqref{s-omega} closest 
to real axis. For weak dephasing, such a pole is given by 
$i\Sigma^{\rm pert}(0)$, yielding additive contributions 
for the total rate $1/T_2^{\rm pert}=\sum_{j=1}^n\alpha_j^2/{t_c}_j$ 
similar to the Matthiessen rule in kinetic theory. 
Equivalently, 
\be \label{coulomb}
{1\over T_2^{\rm pert}} = 
\int \! {d^3\r d^3\r'\over V} \, 
{\la \Omega(\r)\Omega(\r')\ra \over 4\pi D |\r-\r'|} \,.
\ee
(This result can be already deduced from Eq.~\eqref{sigma2} by setting
there $\omega=0$.)

The rate \eqref{coulomb} is formally equivalent to the 
{\it Coulomb energy} of a fictitious charge distribution with 
the charge density $\Omega(\r)$ [or of $4\pi\lambda\Omega_0\chi(\r)$ via locality]. 
Extending this mapping, the variable diffusivity $D(\r)$ would be analogous to 
a variable dielectric constant.
Equivalence of \eqref{coulomb} with Coulomb problem
is due to the Laplace form of Eq.~\eqref{BT}.
In the limit $t\to \infty$,
relaxation is determined by the 
diffusing spins wandering infinitely far to explore the magnetic structure.
The spins then become mediators of the effectively long-ranged 
interaction between the different parts of
the magnetic structure. The long-time asymptote corresponds
to the time-averaged diffusion propagator which becomes a Coulomb 
potential, $\int_0^\infty \! dt \, \Gbare(\r,t) \propto 1/r$.

The self-energy \eqref{sigma2}, as well as the rate \eqref{coulomb}, 
strongly depend on the geometric structure, as the convergence of integrals
is determined by the specific way of how $\Gamma_2$ vanishes at short distances
(large $k$). This nonuniversality can be also seen from 
mapping onto Coulomb problem, since $T_2^{\rm pert}$ maps onto 
capacitance which is sensitive to the conductor geometry.

\subsection{Result for $\ln s(t)$}

The self-energy \eqref{sigma2} substituted into the lineshape \eqref{s-omega} 
is a natural extension of the results \cite{Kiselev2002} 
(rederived in a different 
way in \cite{Sukstanskii2003,Sukstanskii2004_dnr}) 
to the case of arbitrary volume fraction $\zeta$. 
The signal takes the form $s(t) = \exp[-f(t)]$, where 
\be   \label{sexp}
f(\omega) = -{1\over (\omega+i0)^2}
\int \! {d^d\q \over (2\pi)^d}\, 
{{\Gamma_2}(\q) \over -i\omega + D q^2} \,.
\ee
Indeed, while $f(t)\ll 1$, Eq.~\eqref{sexp} is the lowest order 
expansion of \eqref{s-omega} in $\Sigma$, valid for $t\ll t_c/\alpha^2$. 
For $t\gg t_c$, Eq.~\eqref{sexp} yields the same relaxation rate 
as Eqs.~\eqref{s-omega} and \eqref{sigma2}. 
As these two time domains overlap,
the equivalence is proven for all $t$.
Relaxation in the dilute suspension is just a particular case of 
\eqref{sexp}, with $\Gamma_2\simeq\zeta\Gamma_2^{(1)}$, where 
$\Gamma_2^{(1)}$ is the correlator for a single object.

\subsection{Moderate dephasing: 
The self-consistent Born approximation for the lineshape}

Below we attempt to go beyond the perturbative calculation and 
explore the case of moderate dephasing, $\alpha \lesssim 1$.
An exact solution would amount to summing up all the diagrams for the 
self-energy to all orders in $\alpha$, an arduous task even
within a simplifying assumption \eqref{Gamma-delta}. Here we consider the self-energy
in the self-consistent Born approximation (SCBA) 
\be \label{scba}
-\Sigma_{\w,\k} = \int \! {d^d\q\over (2\pi)^d} 
{\Gamma_2(\q)\over -i\w + D(\k+\q)^2 -\Sigma_{\w,\k+\q}}.
\ee
Technically, Eq.~\eqref{scba} amounts to summing all the non-crossing Feynman graphs
(see {\it Methods}, Fig.~\ref{fig:diag}).
The SCBA is akin to mean field theory (more precisely, it 
is equivalent to the mean field treatment of the nonlinear term that 
arises after Gaussian disorder-averaging in the replica field theory). 
Its advantage is that one can move quite far analytically, 
by summing up an infinite subset of Feynman graphs, thereby
collecting contributions from all orders in the external random field $\Omega(\r)$. 
Its disadvantage is that it is uncontrolled 
since it leaves out an infinite subset of graphs for $\Sigma_{\w,\k}$.

Eq.~\eqref{scba} is a complicated integral equation for $\Sigma_{\w,\k}$
due to the $\k$-dependence of $\Sigma$ on the right-hand side.
Since we really only need $\Sigma|_{k=0}$,
we now make another (uncontrolled) simplification that will lead to an 
ansatz for $\Sigma(\w)$. 
Neglecting the $k$-dependence of the self-energy in the denominator,
we obtain the quadratic equation
\be \label{eq-Sigma}
-\Sigma(\w) \simeq {(\delta\Omega)^2 \over -i\w + Dk_c^2 - \Sigma(\w)} 
\ee
whose solution is 
\be \label{Sigma-SCBA}
\Sigma(\w) =   
{1-i\w t_c - R(\w) \over 2t_c} \,, \quad 
R(\w) = \sqrt{(1 -i\w t_c)^2 + 4\alpha^2} . 
\ee
The sign in front of the square root $R(\w)$ in Eq.~\eqref{Sigma-SCBA}
agrees with the perturbative solution \eqref{Sigma2}.

To summarize, both the lowest order signal \eqref{s-pert} and 
the SCBA are perturbative expressions.
However, the lowest order approximation \eqref{s-pert}
is valid for $\alpha \ll 1$, while utilizng the SCBA
allows us to work up until $\alpha \lesssim 1$ (as described below). 
This ``convergence enhancement'',
while questionable mathematically, appears to be quite useful practically,
as we demonstrate below both by comparing the self-energy \eqref{Sigma-SCBA}
to the Monte Carlo simulations, and
by applying it to interpret proton spectra in human blood.

\begin{figure*}[t]
\noindent
{\bf ~~~~~(a)\hspace{\columnwidth} (b)}\\
\includegraphics[width=\columnwidth]{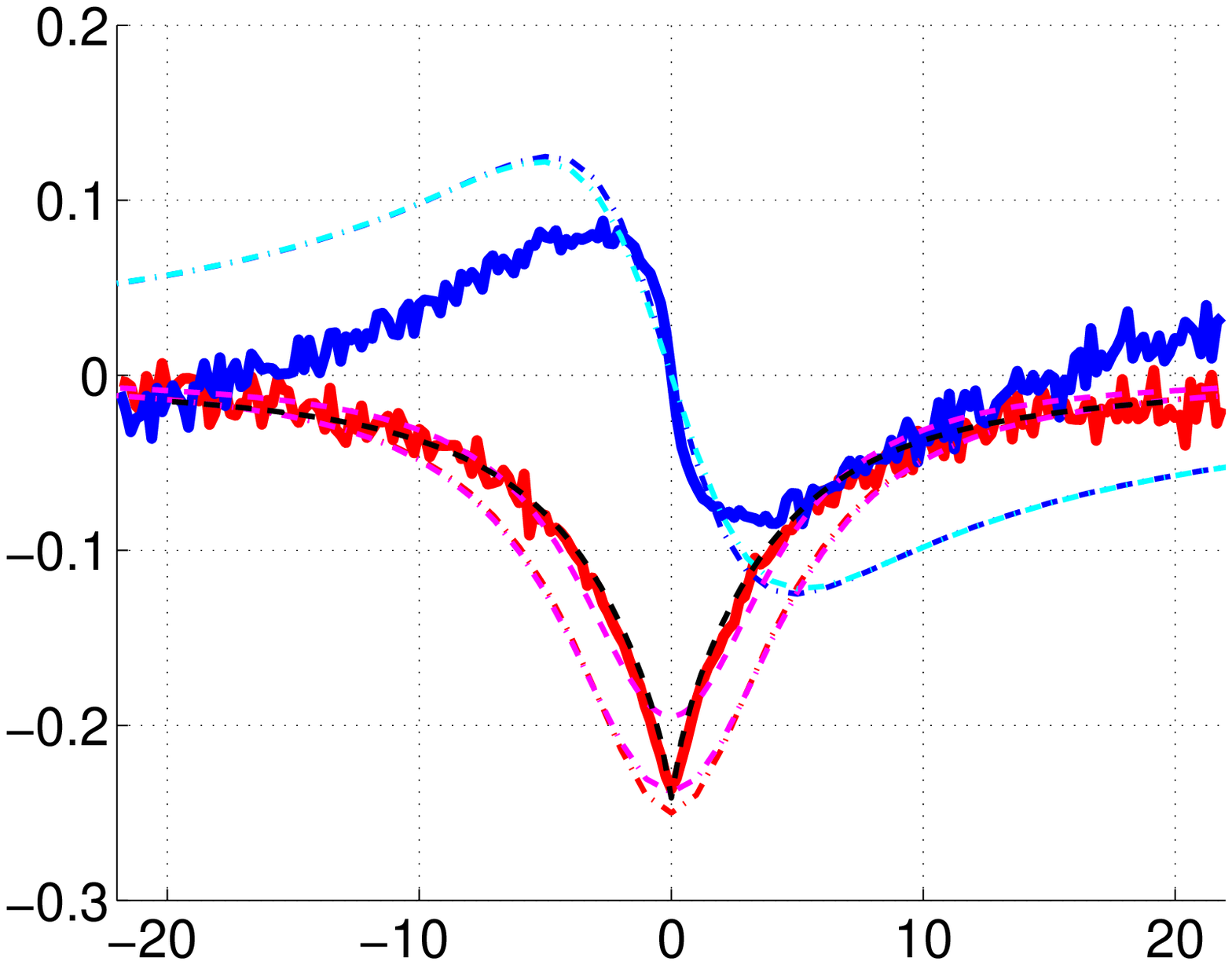}
\includegraphics[width=\columnwidth]{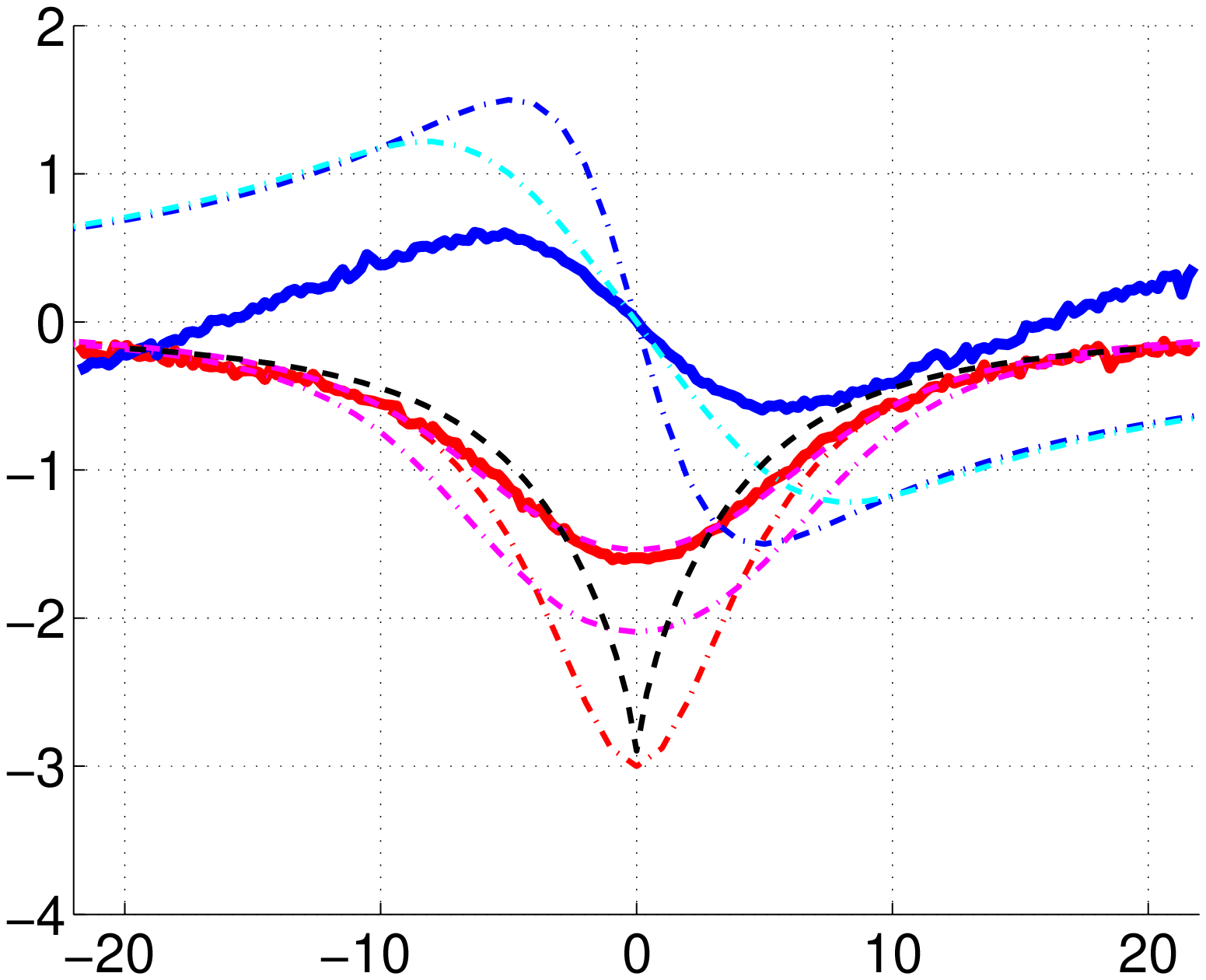}\\
{\bf ~~~~~(c)\hspace{\columnwidth} (d)}\\
\includegraphics[width=\columnwidth]{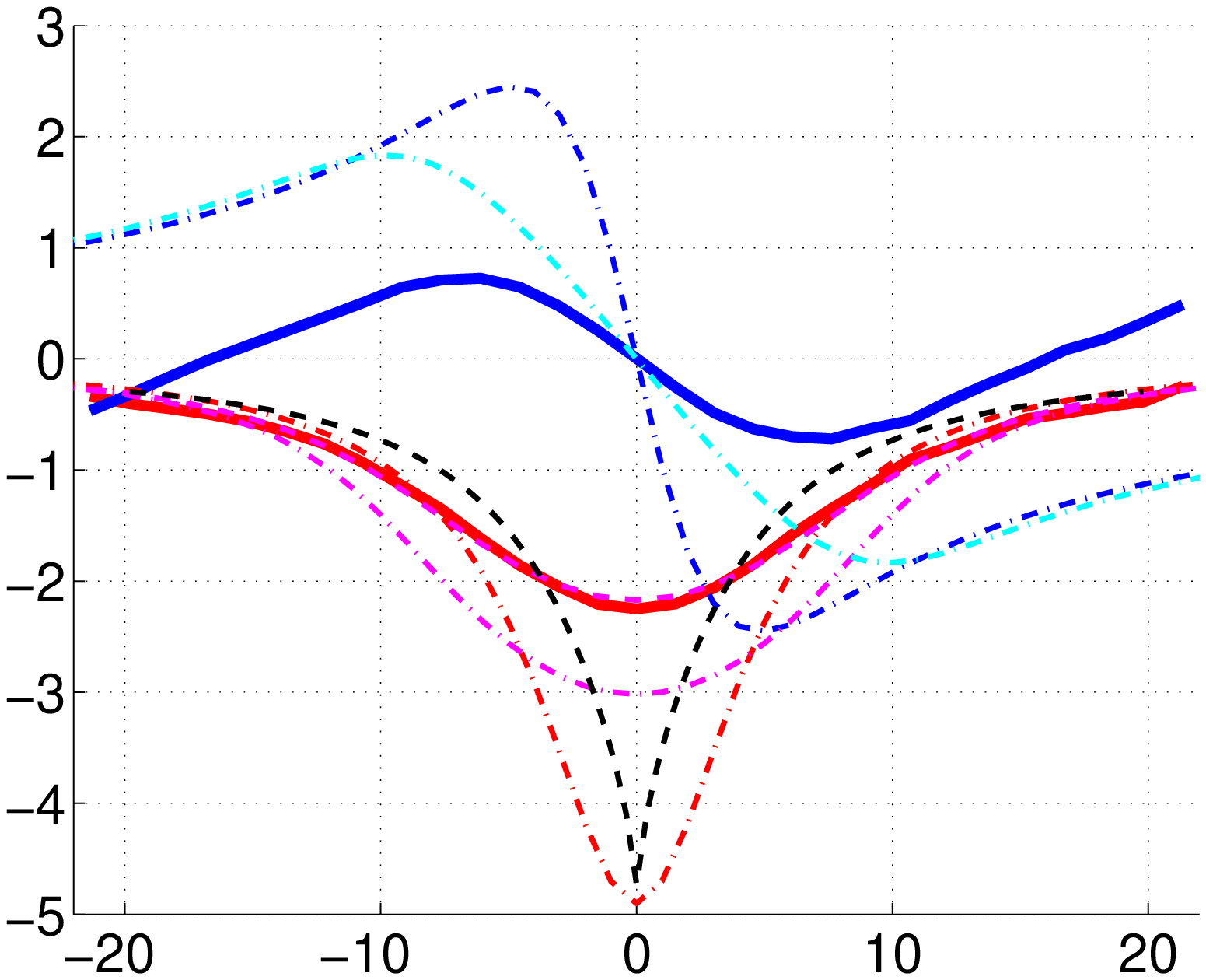}
\includegraphics[width=\columnwidth]{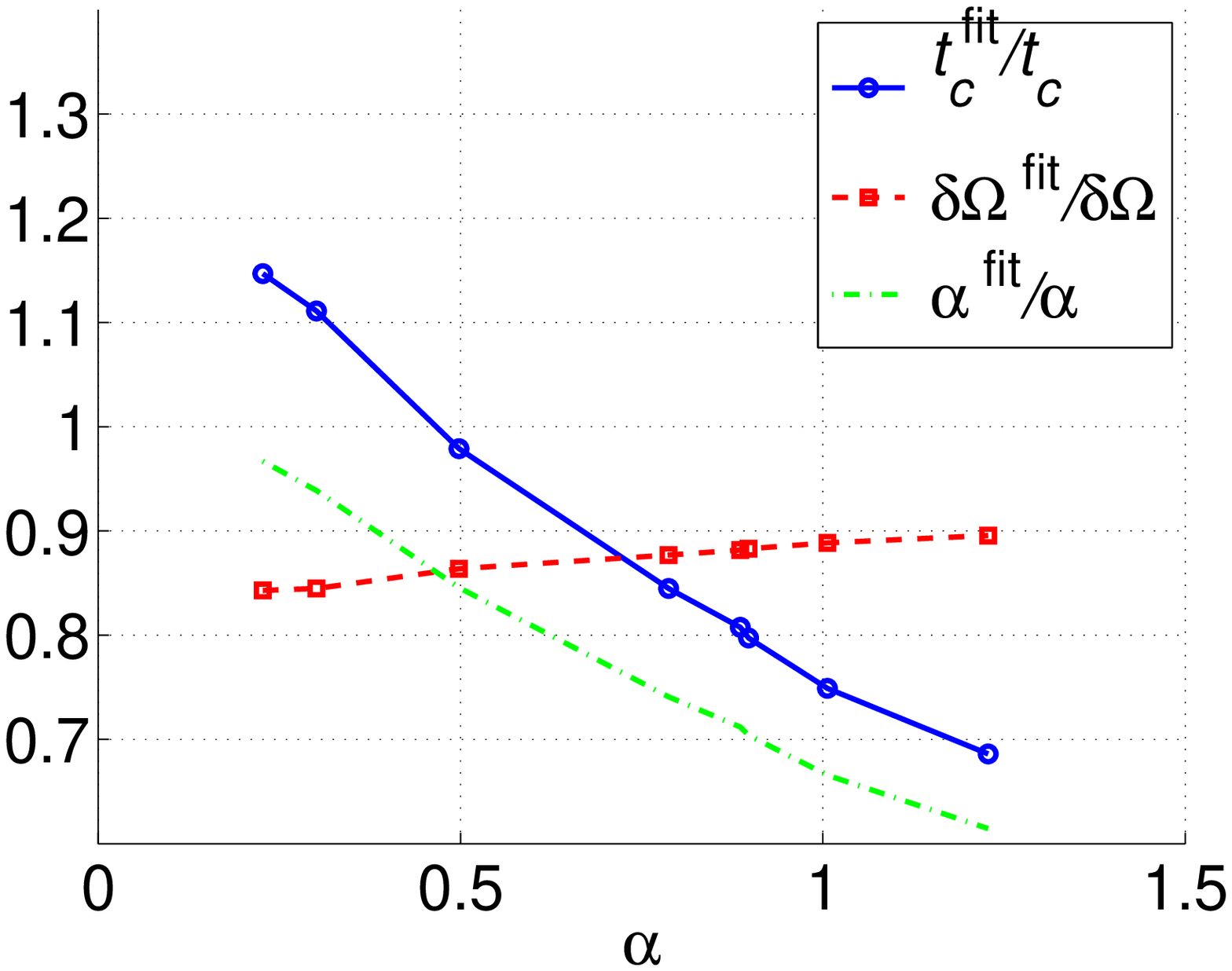}\\

\caption{
Self-energy $\Sigma(\omega)$ obtained in Monte Carlo simulations 
(thick solid lines, red: Re\,$\Sigma$; blue: Im\,$\Sigma$),
compared with the perturbative result for Re\,$\Sigma^{\rm pert}$ 
[Eq.~(\ref{sigma2}) with correlator $\Gamma_2(\q)$ from Fig.~\ref{fig:2d}]
(black dashed).  
Also shown results for the simplified model \eqref{Gamma-delta}
using approximation \eqref{Sigma2} with properly chosen $k_c$ 
[cf. Fig.~\ref{fig:2d}] (dash-dotted red and blue), 
the SCBA \eqref{Sigma-SCBA} (dash-dotted magenta and cyan), 
and the fit of Re\,$\Sigma$ to SCBA ansatz (\ref{Sigma-SCBA}) 
(dashed magenta directly on top of Re\,$\Sigma$), 
with free parameters $t_c$ and $\alpha$.
The values of coupling constant \eqref{alpha} are 
$\alpha=0.23$ (a), $\alpha=0.79$ (b), and $\alpha=1.0$ (c). 
Panel (d) shows the ratio of SCBA fit parameters $t_c^{\rm fit}$,
$\alpha^{\rm fit}$, and $\delta\Omega^{\rm fit}=\alpha^{\rm fit}/t_c^{\rm fit}$   
to their genuine values as function of the true
$\alpha=\delta\Omega\cdot t_c$. 
}
\label{fig:scba_mc}
\end{figure*}

\subsection{Comparison with Monte Carlo simulations}

In Fig.~\ref{fig:scba_mc} we compare the above results 
with Monte Carlo simulation of diffusion and relaxation in the
2d medium described in Fig.~\ref{fig:2d}.
The numerical self-energy 
was calculated according to Eq.~\eqref{s-omega}
after adjusting central frequency of $s(\omega)$ for a small shift $\la \Omega\ra$ 
due to higher-order correlators. 

Practically, the perturbative self-energy \eqref{sigma2}
agrees perfectly with numerics for $\alpha \lesssim 0.3$. 
For these values of $\alpha$ the characteristic ``triangular'' 
shape of $\Re \Sigma(\omega)$ 
deviates from simple Lorentzian (\ref{Sigma2})
due to contribution of harmonics with $k<k_c$. 
Interestingly, for intermediate $\alpha\sim 0.5$, the shape of $\Sigma(\omega)$ 
becomes qualitatively closer to the Lorentzian (\ref{Sigma2}). 
Indeed, for larger $\alpha$, spins dephase before they can explore 
the scales exceeding $l_c$, which increases the relative contribution
of large-$k$ harmonics.

Quantitatively, the SCBA ansatz \eqref{Sigma-SCBA} 
is notably better than both (\ref{sigma2}) and (\ref{Sigma2}) 
for $\alpha\gtrsim 0.5$.
One can attain a perfect agreement of SCBA with the data by
allowing $\alpha$ and $t_c$ to be fitting parameters;
at $\alpha\gtrsim 1$, the SCBA fit values gradually deviate from the geniune 
parameters, 
as illustrated in Fig.~\ref{fig:scba_mc}(d). This is a consequence of 
the properties of the SCBA signal \eqref{s-omega}, \eqref{Sigma-SCBA}
in the complex plane of $z=\omega t_c$: 
When $\alpha>1$, 
the simple pole of \eqref{s-omega} at $z=-i\alpha^2$ dives under the branch cut
connecting $z=-i\pm2\alpha$.
Physically, the developed perturbative approach must break down for 
$\alpha\gtrsim 1$, as the system crosses over from the diffusion-narrowing 
to static dephasing regime, $\alpha\gg 1$. 
The signal in the latter limit coincides with  
the characteristic function of the probability distribution of 
the local Larmor frequency,
$s(t) = \la e^{-i\Omega(\r)t}\ra$. 
The connection of $s(\omega)$ to the mesoscopic 
structure in this limit was studied only for dilute suspensions 
of objects with basic geometries
\cite{Yablonskiy94,Kiselev99,Jensen2000_sdr,Kiselev2001,Sukstanskii2001,Bauer99,Bauer2002}.

\subsection{Comparison with experiment}

We now apply our general results to interpret the line shape of water protons 
in blood as measured by Bj{\o}rnerud {\it et al.} \cite{Bjoernerud2000}. 
Blood plasma was titrated with a superparamagnetic contrast agent
to match the magnetic susceptibility of deoxygenated red blood cells.  
The mesoscopic magnetic structure originated from the susceptibility contrast between plasma and  oxygenated haemoglobin in erythrocytes.

Our model and the expression \eqref{Sigma-SCBA} for the lineshape has been obtained 
assuming unrestricted diffusion (uniform diffusivity $D$). 
Strictly speaking, the assumption of homogeneous diffusivity does not hold for blood.
Below we partially account for the hindered diffusion in blood 
by using the reduced apparent diffusion coefficient (ADC) 
of blood \cite{Kuchel2000}.  
This is a reasonable approximation in view of the fast exchange through the 
cell membrane \cite{Benga87,Waldeck95}.

We have calculated the corresponding self-energy 
from the data of Ref.~\cite{Bjoernerud2000} in three stages: 
First, the imaginary part $\Im s(\w)$ was determined using the 
Kramers-Kr\"onig relations from the published $\Re s(\w)$.
Special attention was paid to phase correction procedure that 
allowed to cancel the residual uniform Larmor frequency offset.
Second, the self-energies $\Sigma_{\rm oxy}$ and $\Sigma_{\rm deoxy}$ 
for the oxygenated and deoxygenated states respectively
were determined according to Eq.~\eqref{s-omega}.
Finally, the difference 
$\Sigma(\w) = \Sigma_{\rm oxy}-\Sigma_{\rm deoxy}$ was formed 
in order to cancel microscopic effects and reduce data processing errors.

\begin{figure}[t]
\includegraphics[width=3.25in]{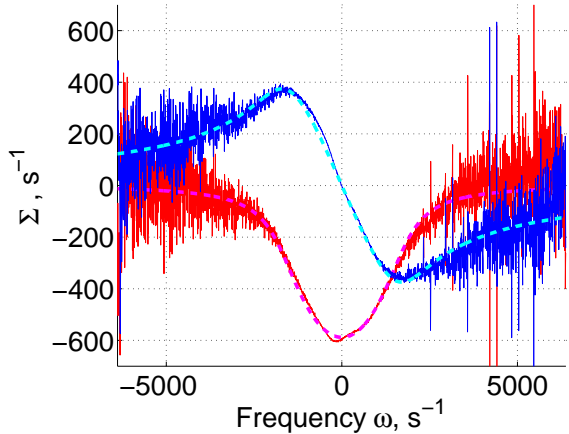}
\caption{
Fit of the human blood spectrum 
from Ref.~\cite{Bjoernerud2000} (solid lines, red:\ $\Re\Sigma$, blue:\ $\Im\Sigma$) 
to the SCBA self-energy \eqref{Sigma-SCBA} (dashed). 
}
\label{fig:scba-fit}
\end{figure}

As demonstrated in Fig.~\ref{fig:scba-fit}, the functional shape \eqref{Sigma-SCBA} 
agrees very well with the self-energy $\Sigma(\omega)$.
The parameters from fitting $\Re \Sigma(\w)$ 
are $\alpha^{\rm fit}=1.3$, $t_c^{\rm fit}=1.5\,$ms, yielding 
$\delta\Omega^{\rm fit}\approx 0.86$\,ms$^{-1}$, and 
$l_c\approx 2.9 \,\mu$m 
assuming the blood ADC value $D=1.1\, \mu{\rm m}^2$/ms at the temperature $T=37\,$C
of the measurement \cite{Bjoernerud2000} 
(we calculated this ADC value from the mean diffusivity of the blood ADC measured at 25\,C \cite{Kuchel2000} asuming Arrhenius temeprature dependence similar to that of water). 
Fitting of Im\,$\Sigma$ gives results within 5\%. 
The value $\delta\Omega^{\rm fit}$ 
is in a reasonable agreement with the experimental $\delta\Omega=0.69$\,ms$^{-1}$. 
The value of $l_c$ matches the size of the doughnut-shaped erythrocytes 
with large diameter $7 \,\mu$m and thickness about $2 \,\mu$m.

\section{Discussion}
\label{sec:discussion}

In this work we suggested the general representation 
\eqref{s-omega} for spectral lineshape $s(\w)$ in the diffusion-narrowing 
regime, in terms of the 
self-energy $\Sigma(\w)$ that contains information about mesoscopic relaxation.
We underscore that, while nominally $s(\w)$ is measured, 
it is the quantity $\Sigma(\w)$ that 
characterizes the mesoscopic medium, in a sense that it is trivial 
when the medium is magnetically homogeneous. We related the self-energy 
dispersion to the structural characteristics of magnetic medium.

The present treatment clearly illustrates the challenges of 
quantifying magnetic media below the spatial resolution: 
The self-avaraging property of the measurement
implies  that two media are equivalent from the point 
of the MR signal if their correlation functions coincide. 
As the sensitivity
to the higher-order correlators $\Gamma_n$ drops fast with increasing $n$,
it is the lowest order $\Gamma_n$, in particular, $n=2$, that are most important.
Since there are an infinite variety of media whose lowest order $\Gamma_n$ coincide,
the inverse problem is, strictly speaking, unsolvable. 
As with any ill-posed inverse problem, having {prior knowledge} about
the system is crucial. In particular,
knowledge about the number of characteristic
length scales in the geometric profile of susceptibility allows one to select
the minimal number of the basis functions \eqref{Gamma-delta-r} and 
to construct the corresponding SCBA ansatz by generalizing the form
(\ref{Sigma-SCBA}). 
Applying the proposed approach in biomedical MRI may allow one to
quantify biophysical tissue properties that can be further related 
to physiological processes and malfunctions.

We conclude by noting that the present approach based on the lineshape
 \eqref{s-omega}
and on the simple form of the correlator \eqref{Gamma-delta} is completely general.
With straightforward modifications, it can be applied to ``resolve'' the mesoscopic 
details in diffusion, in conductivity, and in light propagation
in heterogeneous condensed matter systems.

\section*{Acknowledgments}
It is a pleasure to thank Atle Bj{\o}rnerud for providing us 
with the spectroscopic data.
D.N. was supported by NSF grants No. DMR-0749220 and No. DMR-0754613.

\begin{figure}[t]
{\bf (a)}

\includegraphics{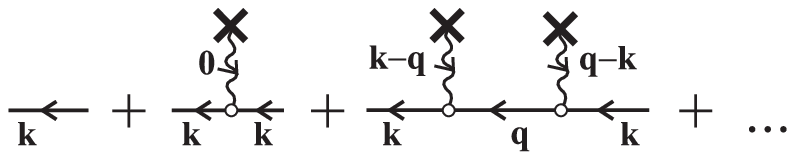}
\vspace{0.5cm}

{\bf (b)}

\includegraphics{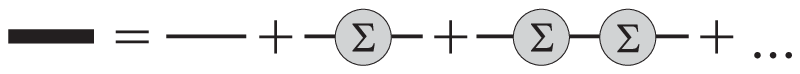}

{\bf (c)}

\includegraphics{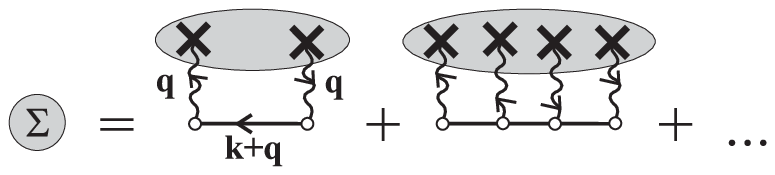}

{\bf (d)}

\includegraphics{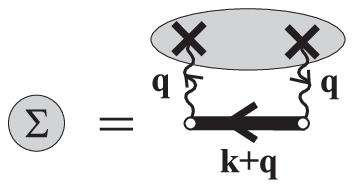}
\caption{
(a) The symbolic representation of the Born series \eqref{Born}.
Wavy lines represent interaction with the static heterogeneous 
Larmor frequency offset $-i\Omega_\k$.
Solid lines represent the free propagators $\Gbare_{\omega,\k}$
\eqref{Gbare}.
(b) The series \eqref{sigma-series} for the ensemble-averaged propagator
in terms of the self-energy.
(c) Perturbative contributions to the self-energy. 
The first diagram is given by Eq.~\eqref{sigma2}.
(d) Symbolic form of the SCBA Eq.~\eqref{scba}.
}
\label{fig:diag}
\end{figure}

\appendix
\section{Appendix: Methods}
\label{sec:methods}

The calculation of the averaged propagator $G$ is done in two stages:
(i) finding the propagator $\G(\r,\r_0;t)$ of Eq.~\eqref{BT}, 
and (ii) averaging over the realizations of $\Omega(\r)$.
Below we describe these stages making use of symbolic notation 
adopted from quantum theory \cite{AGD}.

On the stage (i), one uses the fact the exact Green's function 
$\G \equiv \L^{-1}$ is the inverse of the Bloch-Torrey differential operator 
\be \label{L}
\L = \partial_t - D\nabla^2 - U\,, \quad U = -i\Omega \,,
\ee
or, equivalently, $\L \G = \delta(\r-\r_0)\delta(t)$.
We now define the bare propagator $\Gbare = \L_0^{-1}$ as the 
fundamental solution of the diffusion equation,
\be \label{L0}
\L_0  \Gbare = \delta(\r-\r_0)\delta(t) \,, 
\quad \L_0 = \partial_t - D\nabla^2 \,.
\ee
The function $\Gbare$, whose Fourier transform is Eq.~\eqref{Gbare}, 
has a familiar form in three dimensions,
\be \label{Gbare-rt}
\Gbare(\r;t) = \theta(t) (4\pi D t)^{-3/2} e^{-r^2/4Dt}
\ee
[where $\theta(t>0)=1$ and $\theta(t<0)=0$ is the step function].
The exact Green's function $\G$ is then obtained by the summation of the 
operator geometric series 
\bea \non
\G &=& (\L_0-U)^{-1}  
= (1-\Gbare * U)^{-1} * \Gbare \\
&=& \Gbare + \Gbare * U * \Gbare 
+ \Gbare * U * \Gbare * U * \Gbare \dots
\label{Born}
\eea
The latter series is analogous to the Born series for the scattering 
amplitude in quantum theory \cite{Landau3,AGD}.
The Born series can be schematically represented by the sum of the Feynman graphs 
(Fig.~\ref{fig:diag}). 

The distribution-averaging stage (ii)
formally amounts to substituting the products of 
$\Omega(\r_1)...\Omega(\r_n)$ by the corresponding 
correlators $\Gamma_n$.
We denote this by joining the crosses in the Feynman diagrams
(Fig.~\ref{fig:diag}) into all possible combinations 
symbolizing $\Gamma_n$.

The exact averaged propagator in Eq.~\eqref{s-G} can be represented in terms of
the {self-energy} $\Sigma_{\omega, \k}$
which is a sum of all {irreducible} contributions to $G$
(by irreducible diagram we mean the one which cannot be cut into two 
by removing any internal solid line) \cite{AGD}. 
The self-energy $\Sigma$ can be then used for the block summation 
[Fig.~\ref{fig:diag}(b)] to obtain the distribution-averaged Green's function 
\be \label{sigma-series}
G_{\omega,\k} = \Gbare_{\omega,\k} 
+ \Gbare_{\omega,\k} \Sigma_{\omega,\k} \Gbare_{\omega,\k} 
+ \dots \ ,
\ee
which is equivalent to Eq.~\eqref{G}.

We end this part by outlining 
the properties of the exact Green's function \eqref{G}.
First, we note that the
magnetization conservation for short times, $s(0)=1$,
fixes the normalization for large frequency behavior
irrespectively of the medium: 
\be \label{sumrule}
G_{\omega, \k}|_{\omega\to\infty} = -{1\over i\omega} \,.
\ee
Second, due to causality, as for any response function, 
$G(t,\r)|_{t<0} \equiv 0$, which requires 
that $G_{\omega,\k}$ [and thereby $s(\omega)$] be analytic in the upper-half-plane 
of the complex variable $\omega$.



\end{document}